\documentclass[twocolumn,showpacs]{revtex4}

\usepackage{graphicx}%
\usepackage{dcolumn}
\usepackage{amsmath}

\makeatletter
\def\btt#1{\texttt{\@backslashchar#1}}%
\DeclareRobustCommand\bblash{\btt{\@backslashchar}}%
\makeatother

\begin{document}

\title[Short Title]{Anomalous diffusion, nonlinear fractional Fokker-Planck equation and solutions}
\author{E. K. Lenzi$^{1,2}$, L. C. Malacarne$^{2}$, R. S. Mendes$^{2}$ and I. T. Pedron$^{2}$}
\affiliation{ $^{1}$Centro Brasileiro de Pesquisas F\'\i sicas, R.
Dr.
Xavier Sigaud 150, 22290-180 Rio de Janeiro-RJ, Brazil \\
$^{2}$Departamento de F\'\i sica, Universidade Estadual de
Maring\'a, Av. Colombo 5790, 87020-900 Maring\'a-PR, Brazil}

\date{\today}
\begin{abstract}
We obtain new exact classes of solutions for the nonlinear
fractional Fokker-Planck-like equation $\partial_t \rho=
\partial_x \left \{ D(x) \partial^{\mu -1}_x \rho^{\nu} - F(x)
\rho \right \} $ by considering a diffusion coefficient $D =
{\cal{D}} |x|^{-\theta}$ ($\theta \in {\cal R}$ and ${\cal{D}}>0$)
and a drift force $F= -k_{1}x+ \overline{k}_{\gamma}
x|x|^{\gamma-1}$ ($k_{1}, \overline{k}_{\gamma}, \gamma \in {\cal
R}$). Connection with nonextensive statistical mechanics based on
Tsallis entropy is also discussed.
\end{abstract}
\keywords{Tsallis entropy, anomalous diffusion }

 \pacs{PACS numbers: 82.20.Db,66.10.Cb,05.60.+w, 05.40.+j}

 \maketitle

\section{Introduction}

The nonlinear or fractional Fokker-Planck-like equations
\cite{shlesinger,fractional,goldenfeld,T88,hilfer} have been used
to analyze several physical situations that present anomalous
diffusion. A typical  nonlinear diffusion equation is $\partial_t
\rho\,=\,{\cal{D}} \partial_{xx}^{2} \rho^\nu$, usually
denominated as {\it porous medium equation}. It has been employed
in the analysis of percolation of gases through porous media
($\nu\geq 2$) \cite{buc6}, thin saturated regions  in porous media
($\nu=2$) \cite{buc7}, a standard solid-on-solid model for surface
growth ($\nu=3$), thin liquid films spreading under gravity
($\nu=4$) \cite{buc8}, among others \cite{buc10}. In the presence
of an external drift force, a drift term is incorporated to the
porous medium equation. For example, such situation has been
considered in the study of escape time, or mean first passage
time, by considering a nonlinear Fokker-Planck-like equation,
leading eventually to a generalization of the Arrhenius law
\cite{celia}. On the other hand, the fractional diffusion
equations are frequently considered as alternative approaches to
continuous time random walk models, generalized Langevin
equations, or generalized master equations. For instance, we can
mention that a generalized master equation, related to a
fractional Fokker-Planck equation, has been used in the modelling
of nonmarkovian dynamical processes in protein folding
\cite{proteinfolding}, describe relaxation to equilibrium in
system (such as polymers chains and membranes) with long temporal
memory \cite{relaxacao}, and in anomalous transport in disordered
systems\cite{anomaloustransport}. Another class of anomalous
diffusion can be obtained using a spatial dependent diffusion
coefficient $D=D(x)$. For instance, in \cite{prl} $D\propto
|x|^{-\theta}$ has been employed to analyse the diffusion in a
fractal medium.

Typical  physical situations where these kinds of equations are
 applied  concern  anomalous diffusion of the correlated type
(both sub- and superdiffusion; see \cite{correlacionada} and
references therein) and of the L\'evy type (superdiffusion; see
\cite{levy} and references therein).  The anomalous correlated
diffusion, when it has a finite second moment $\langle x^2 \rangle
\propto t^{\sigma}$,   corresponds to superdiffusion, normal
diffusion and subdiffusion for $\sigma > 1$, $\sigma=1$ and $0 <
\sigma < 1$, respectively; $\sigma =0$ basically indicates a
localization. For L\'evy diffusion ( L\'evy distributions), there
is  no finite second moment, i.e., $\langle x^2 \rangle$ diverges.

We dedicate the present work to investigate the features that
emerge from considering a mixing between fractional derivative,
nonlinear aspects and spatial dependent diffusion coefficient.
More precisely, we focus the nonlinear fractional
Fokker-Planck-like equation:
\begin{equation}
\frac{\partial}{\partial t}\rho(x,t)=\frac{\partial}{ \partial
x}\left \{ D(x) \frac{\partial^{\mu -1}}{\partial x^{\mu -1}}
[\rho(x,t)]^\nu - F(x) \rho(x,t) \right\} \label{eq-1},
\end{equation}
where $\nu,\mu \in {\bf {\cal{R}}}$, $D(x) \propto |x|^{-\theta}$
is a (dimensionless) diffusion coefficient ($\theta \in {\bf
{\cal{R}}})$, and $F(x) \equiv -dV(x)/dx$ is a (dimensionless)
external force ({\it drift}) associated with the potential $V(x)$.
We use the Riemann-Liouville operator\cite{Oldham} for the
fractional derivative, and we also work with the {\it positive}
$x$ axis and, later on, we employ symmetry to extend the results
to the entire real axis (in other words, we are working with
$\partial^{\mu-1}/\partial |x|^{\mu-1}$). In addition, the initial
condition $\rho(x,0)=\delta(x)$ and the boundary condition
$\rho(\pm \infty,t) \rightarrow 0$ are used. Notice that the
normalization of $\rho$ is time independent. Indeed, if we write
the equation in the form $\partial _t \rho =
\partial _x {\cal {J}} $ and assume the boundary conditions ${\cal
{J}}(\pm \infty,t) \rightarrow 0$, it can be shown that
$\int_{-\infty}^{\infty} dx \;\rho(x,t)$ is a constant of motion.

As indicated above, Eq.(\ref{eq-1}) can be used to describe a
large class of anomalous diffusion processes since it contains, as
a particular case, the porous medium equation, L\'evy
superdiffusion, as well as a mix of them. For $\mu=2$ and $\nu=1$,
Eq.(\ref{eq-1}) recovers the standard Fokker-Planck equation with
a drift term. The particular case $F(x)=0$ (no drift), $D(x)=
constant$, $\mu=2$,  and arbitrary $\nu$ has been considered by
Spohn \cite{spohn}. The case $\mu =2$ has also been considered in
\cite{plastinos,buckman,isabel} with simple drifts. The case $\mu
<2$ with $F(x)=0$ and $D=constant$ has been addressed in
\cite{ct2000}. Our present analysis involves several extensions of
these cases, in particular, by employing the external force
$F=-k_{1}x +\overline{k}_{\gamma}x |x|^{\gamma-1}$. Indeed, a
variety of systems, which presents anomalous diffusion, can be
described by short or long-range forces when the previous one is
employed. This analysis is presented in Sec. II for the case
$\mu=2$; the solution is obtained in a closed form. In Sec. III,
we also obtain exact results for the case $\mu \neq 2$, i.e., for
the external drift $F=-k_{1}x$ we obtain solutions for $\mu \in
{\cal R}$, and for $F=-k_{1}x+k_{\gamma}x|x|^{\gamma-1}$ we find
solutions for $\mu=0$ and $\mu=1$. Finally, in Sec. IV we present
our conclusions.


\section{ Nonlinear Fokker-Planck equation}

Before starting our discussion, it is interesting to note that the
results presented in
\cite{plastinos,buckman,isabel,ct2000,duxbury2,renio} may be
essentially obtained by using normalized scaled solutions of the
type
\begin{equation}
\label{eq-2} \rho\left(x,t \right)=\frac{1}{\Phi \left(t \right)}
\tilde{\rho}\left[\frac{x}{\Phi \left(t \right)} \right]
\end{equation}
in Eq.(\ref{eq-1}). Similar approach has been employed, for
instance, in \cite{goldenfeld,duxbury2}. To illustrate this
procedure let us insert  Eq.(\ref{eq-2}) into Eq.(\ref{eq-1}) with
$\mu=2$, $F(x)=0$ and $D(x)={\cal {D}}=constant$. Thus,
\begin{equation}
\label{eq-3}  -\frac{\dot{\Phi} \left(t \right)}{{\Phi} \left(t
\right)^{2}}\frac{d}{dz}[ z\tilde{\rho}\left(z
\right)]=\frac{d}{dz} \left\{ \frac{{\cal {D}}}{\Phi \left(t
\right)^{2+\nu}} \frac{d}{dz}\left[\tilde{\rho}\left(z \right)
\right]^{\nu } \right\}\;\; .
\end{equation}
In the following, we eliminate the explicitly time dependence on
Eq. (\ref{eq-3})  choosing
\begin{equation}\label{eq-3b}
[\Phi(t)]^{\nu} \frac{d}{dt} \Phi(t)=k  \;,
\end{equation}
where $k$ is a constant. Thus, Eq. (\ref{eq-3b}) replaced into Eq.
(\ref{eq-3}) gives us
\begin{eqnarray}
\label{eq-3c} -k\frac{d}{dz} z\tilde{\rho}\left(z \right)=
\frac{d}{dz} \left \{{\cal {D}}
\frac{d}{dz}\left[\tilde{\rho}\left(z \right) \right]^{\nu }
 \right \}
 \;\; .
\end{eqnarray}
From Eq. (\ref{eq-3b}) we obtain
\begin{equation}
 \Phi(t)=\left[(1+\nu)k t\right]^{1/(1+\nu)}~,
\end{equation}
where we have adopted the solution which satisfies $\Phi(0)=0$ for
$\nu>-1$. Finally, the solution of Eq. (\ref{eq-3c}) leads to
\begin{eqnarray}
\label{eq-4} \rho(x,t)= \frac{1}{\Phi(t)} \exp_q \left [
-\frac{k}{2 {\cal {D}}\nu} \left(\frac{x}{\Phi(t)} \right)^2
\right] \;\; ,
\end{eqnarray}
with $q=2-\nu$ and $\exp _{q}(x) \equiv
[1+(1-q)x]^{\frac{1}{1-q}}$ being the $q$-exponential function.
This generalized exponential  arises within the nonextensive
thermostatistical formalism by optimizing, under appropriate
constraints, the entropic form $S_q=(1- \int dx
[\rho(x,t)]^q)/(q-1)$ \cite{T88,tsallisnonextensive}. The constant
$k$ can be fixed  from the normalization condition
$\int_{-\infty}^{\infty} dx \; \rho(x,t)=1$.

Let us now go back to Eq. (\ref{eq-1}) with
$F=-k_{1}x+\overline{k}_{\gamma}x|x|^{\gamma-1}$ ($k_{1}\neq
\overline{k}_{\gamma}$) and $D(x)={ \cal { D}}|x|^{-\theta}$
($\theta \in {\cal R}$). Firstly, we address our discussion to
case $\mu=2$.  In this case, Eq.(\ref{eq-1}) reduces to
\begin{widetext}
\begin{eqnarray}
\label{eq-s1} -\frac{\dot{\Phi} \left(t \right)}{{\Phi} \left(t
\right)^{2}}\frac{d}{dz} [z\tilde{\rho}\left(z \right)]=
\frac{d}{dz} \left \{ \frac{z^{-\theta}{\cal {D}} }{\Phi \left(t
\right)^{2+\nu+\theta}} \frac{d}{dz}\left[\tilde{\rho}\left(z
\right) \right]^{\nu }+ \left [ \frac{zk_{1}}{\Phi(t)}-\frac
{z^{\gamma} \overline{k}_{\gamma}} {\Phi(t)^{2-\gamma}} \right ]
\tilde{\rho}(z) \right \}
\end{eqnarray}
when $\rho(x,t)$, given by Eq. (\ref{eq-2}), is employed.  We are
interested in situations for which the scaled solution of the type
indicated in Eq. (\ref{eq-2}) is still valid, i.e., we would like
to uncouple the $t$ and $z$ dependence in Eq. (\ref{eq-s1}) as
performed in Eq. (\ref{eq-3}) via Eq. (\ref{eq-3b}). This can be
accomplished when  $\gamma + \theta + \nu=0$. If this condition is
satisfied, we obtain
\begin{eqnarray}
\label{eq-s2} \frac{\dot{\Phi} \left(t \right)}{\Phi \left(t
\right)^{2}}+ \frac{k_{1}}{\Phi(t)}=\frac{k'}{\Phi \left(t
\right)^{2+\nu+\theta}}
\end{eqnarray}
and
\begin{eqnarray}
\label{eq-s3} -k'\frac{d}{dz} z\tilde{\rho}\left(z \right)=
\frac{d}{dz} \left \{{\cal {D}} z^{-\theta}
\frac{d}{dz}\left[\tilde{\rho}\left(z \right) \right]^{\nu }
-k_{\gamma}z^{\gamma} \tilde{\rho}(z) \right \}
 \;\; ,
\end{eqnarray}
where $k^{\prime}$ is a constant that plays a role analogous to
$k$ in Eq. (4), and it is to be determined through the
normalization.

From the solutions of Eq. (\ref{eq-s2}) and Eq. (\ref{eq-s3}),
which generalize the results of the previous example, we have
\begin{eqnarray}\label{eq-s4b}
\Phi(t)=\left[\frac{k^{\prime}}{k_{1}} \left(
1-e^{-(1+\nu+\theta)k_{1}t} \right) \right]^{1/(1+\nu+\theta)}
\;\;
\end{eqnarray}
and
\begin{eqnarray}\label{eq-8}
\rho(x,t)=  \frac{1}{\Phi(t)} \exp_q \left[-\frac{1}{{\cal
{D}}\nu} \left( \frac{k'}{2+\theta}
\left(\frac{|x|}{\Phi(t)}\right)^{2+\theta} -k_{\gamma}
\ln_{2-q}\left( \frac{|x|}{\Phi(t)} \right) \right)\right] \;,
\end{eqnarray}
\end{widetext}
 where $\ln_q x \equiv (x^{1-q}-1)/(1-q)$ is the
$q$-logarithm function (that is the inverse function of the
$q$-exponential).

In the previous study for the case $\mu=2$, we notice that the
analytical solution (\ref{eq-8}) takes  fractal and nonlinear
aspects into account, since there is a spatial dependence on the
diffusion coefficient $(\theta \neq 0)$ and a nonlinear term $(\nu
\neq 1)$. In fact, it contains  results obtained in
\cite{prl,spohn,plastinos,buckman,isabel,duxbury2,renio,tsalliseklenzi}
as particular case. Moreover, it reduces to the Rayleigh
process\cite{strato} case in the limit $q\rightarrow 1$ and
$\theta = 0$.


\section{ Nonlinear fractional Fokker-Planck equation}

In order to give an extension for the previous  discussion
($\mu=2$), we address our investigation concerning Eq.
(\ref{eq-1}) by considering $\mu \neq 2$. Without loss of
generality we are considering ${\cal{D}}=1$. We follow the
procedure employed in \cite{duxbury2,tsalliseklenzi} taking  the
generic property
\begin{equation}
\label{eq-10} \frac{d^{\delta  }}{dx^{\delta }}{\cal{G}}\left(
ax\right)= a^{\delta }\frac{d^{\delta  }}{d\overline{z}^{\delta
}}{\cal{G}}\left( \overline{z}\right)\;\;\;\;(\delta \in {\cal R})
\end{equation}
with $ \overline{z}=ax$ into account. This basic property holds
not only for the ordinary derivative but also for all fractional
operators that we are considering here. Thus, substituting Eq.
(\ref{eq-2}) into Eq. (\ref{eq-1}) and imposing
\begin{equation}
\label{eq-11} \frac{\dot{\Phi} \left(t \right)}{\Phi \left(t
\right)^{2}}+ \frac{k_{1}}{\Phi(t)}=-\frac{\overline{k}'}{\Phi
\left(t \right)^{\nu+\theta+\mu}} \;\; ,
\end{equation}
in order to uncouple $t$ and $z$ dependence, where $\overline{k}'$
is an arbitrary constant. Therefore, we obtain
\begin{equation}
\label{eq-12} \Phi \left(t
\right)=\left[\overline{k}_{2}-\frac{\overline{k}'}{k_{1}}
\left(1-e^{-(\mu+\nu+\theta-1)k_{1}t} \right)
\right]^{\frac{1}{\nu+\mu+\theta - 1}}
\end{equation}
and
\begin{equation}
\overline{k}'\frac{d}{dz}\left[z \tilde{\rho}\left(z \right)
\right] =\label{eq-13} \frac{d}{dz} \left \{
z^{-\theta}\frac{d^{\mu - 1}}{dz^{\mu - 1}}\left[
\tilde{\rho}\left(z \right) \right] ^{\nu } -k_{\gamma}
z^{-\theta-\nu -\mu+2} \tilde{\rho}(z) \right \}\;\; ,
\end{equation}
with $\gamma=-\theta-\nu -\mu+2$ and $\overline{k}_2$ being an
arbitrary constant. If $\overline{k}_2=0$ we have $\Phi (0)=0$ for
$\nu+\mu+\theta >1$, recovering Eq. (\ref{eq-s4b}) when $\mu
\rightarrow 2$. By considering $\overline{k}_2= 0$ corresponds
$\rho(x,0)\propto \delta (x)$. On the other hand, employing
$\overline{k}_2\neq 0$ leads to $\rho(x,0)$  less concentrated
than $\delta (x)$. A more tractable equation to $\tilde{\rho}(z)$
is obtained after an integration of Eq. (\ref{eq-13}), i. e.,
\begin{equation}
\label{eq-14} \overline{k}' z \tilde{\rho}(z) =
z^{-\theta}\frac{d^{\mu - 1}}{dz^{\mu - 1}}\left[ \tilde{\rho}(z)
\right]^{\nu }-k_{\gamma}z^{-\nu-\theta-\mu+2} \tilde{\rho}(z)
+{\cal C} \;\; ,
\end{equation}
where ${\cal C}$ is another arbitrary constant. In the following,
we address our discussion focusing two situations of
Eq.(\ref{eq-14}). The first one is characterized by a linear
drift, $F=-k_{1}x$. In the second one, we consider the drift
$F=-k_{1}x+k_{\gamma}x|x|^{\gamma-1}$ for particular values of
$\mu$.

\subsection{The drift $F=-k_{1} x$}

Let us start our discussion by considering Eq.(\ref{eq-14}) with $k_{1}\neq 0$ and
$k_{\gamma}=0$. To solve it, we can use the procedure employed in \cite{ct2000}
with ${\cal C}=0$ and
\begin{eqnarray}
\nu &=&\frac{2-\mu}{1+\mu+\theta} \;\;.
\end{eqnarray}
In this case, we have
\begin{eqnarray} \label{eq-18} \rho\left(x,t
\right) =\frac{A}{\Phi(t)}
\left[\frac{z^{(\mu+\theta)(1+\mu+\theta)}} {\left(1+b z
\right)^{(1-\mu)(1+\mu+\theta)}}\right]^{\frac{1}{1-2\mu-\theta}}
\end{eqnarray}
with $\Phi(t)$ given by Eq.(\ref{eq-12}), $z\equiv x/\Phi(t)$, and
\begin{eqnarray}
A=\left[\overline{k}'\frac{\Gamma\left(-\beta\right)}
{\Gamma\left(\alpha+1\right)}\right]^{\frac{1+\mu+\theta}{1-2
\mu-\theta}}.
\end{eqnarray}
Note that Eq.(\ref{eq-18}) reduces that one obtained in
\cite{tsalliseklenzi} in the absence of drift. We notice that the
presence of source (absorvent) term, $\alpha \rho$ $(\alpha =
const)$, in the right hand of Eq.(\ref{eq-1}) modifies
Eq.(\ref{eq-12}) as follows:
\begin{widetext}
\begin{eqnarray}
\Phi(t)=\left[\overline{k}_2-\frac{(\mu+\theta+\nu-1)\overline{k}'}{(\nu-1)\alpha+(\mu+\theta+\nu-1)
k_{1}}\left(e^{(\nu-1)\alpha
t}-e^{-(\mu+\theta+\nu-1)k_{1}t}\right) \right]^{\frac{1}{\mu
+\theta+\nu-1}}\; .
\end{eqnarray}

\subsection{The drift $ F=-k_{1}x + k_{\gamma} x|x|^{\gamma-1}$}

Now, we consider other particular cases, namely $\mu = 0$ and $\mu
= 1$. Let us start with $\mu = 0$ and arbitrary $\nu$. The
corresponding equation is

\begin{equation}\label{-1}
\frac{\partial }{\partial t}\rho\left( x,t \right)= \frac{\partial
}{\partial x}\left \{ x^{-\theta} \int_{0}^{x}\left[\rho\left(y,t
\right)\right]^{\nu}dy + \left [ k_{1} x - \overline{k}_{\gamma}
x^{-\nu-\theta+2} \right ] \rho(x,t) \right \} \; .
\end{equation}
To solve it  let us go back to Eq. (\ref{eq-14}) with $\mu=0$ and
${\cal C}=0$, i.e.,
\begin{equation}
\label{boh} \overline{k} z\tilde{\rho}(z)=z^{-\theta} \int_{0}^{z}
d\overline{z}\left[\tilde{\rho}\left(\overline{z}\right)\right]^{\nu}-\overline{k}_{\gamma}
z^{-\nu-\theta+2} \tilde{\rho}(z) \;,
\end{equation}
whose solution is given by
\begin{eqnarray}
\tilde{\rho}(z) \propto \frac{1}{\overline{k}'z^{1+\theta}+
\overline{k}_{\gamma}z^{-\nu+2}}\left[1+\tilde{{\cal{C}}} \int
dz\left( \overline{k}'z^{1+\theta}
+\overline{k}_{\gamma}z^{-\nu+2} \right)^{-\nu}
\right]^{1/(1-\nu)} \;,
\end{eqnarray}
where $\tilde{{\cal{C}}}$ is a constant.

Let us now address our analysis to the case $\mu = 1$. It
corresponds to investigate the equation
\begin{equation}
\label{zero} \frac{\partial }{\partial t}\rho\left( x,t \right)=
\left. \left. \frac{\partial}{\partial x} \right \{
x^{-\theta}\left[\rho\left(x,t \right) \right] ^{\nu}+\left[
k_{1}x-\overline{k}_{\gamma}x^{-\nu-\theta+1} \right]\rho(x,t)
\right \} \;\;.
\end{equation}
\end{widetext}

To obtain the solution of this equation employing the $ansatz$
(\ref{eq-2}) we use Eq.(\ref{eq-14}). It follows that
\begin{equation}
\label{gam0} \overline{k}'z\tilde{\rho}\left(z
\right)=z^{-\theta}\left[ \tilde{\rho}(z)
\right]^{\nu}-\overline{k}_{\gamma}
z^{-\nu-\theta+1}\tilde{\rho}(z)+{\cal C} \; .
\end{equation}
The solution corresponding to ${\cal C}=0$ is
\begin{equation}
\tilde{\rho} \left(z\right) \propto
\left(\overline{k}'z^{1+\theta}+
\overline{k}_{\gamma}z^{-\nu+1}\right)^{1/(\nu-1)} \; .
\end{equation}

Let us finally indicate a connection between the results obtained
here and the solutions that arise from the optimization of the
nonextensive entropy\cite{T88}. These distributions do not
coincide for arbitrary value of $x$. However, the comparison
between the asymptotic behaviors (large $|x|$) enables us to
identify the type of tails. For instance, by taking the behavior
exhibited in Eq.(\ref{eq-18}) for large $|x|$ with the asymptotic
behavior $1/|x|^{2/(q-1)}$ into account, which appears in
\cite{T88} for the entropic problem, we obtain
\begin{eqnarray}
q=\frac{3+\mu+\theta} {1+\mu+\theta} \;\; .
\end{eqnarray}
This relation is the same obtained in \cite{tsalliseklenzi} in the
absence of drift.  It also recovers that one already established
in \cite{ct2000} for $\theta =0$.


\section{Conclusions}

In summary, we have considered the one-dimensional nonlinear
fractional Fokker-Planck equation (Eq.(\ref{eq-1}) for some
space-dependent (power-law) classes of external drift and
diffusion coefficient. We have shown that it admits exact
solutions where space scales with a (usually simple) function of
time (as indicated in Eq.(\ref{eq-2})), and have fully and
explicitly worked out some of those. In particular, we first
consider the case $\mu=2$ in the presence of the drift,
$F=-k_{1}x+\overline{k}_{\gamma}x|x|^{\gamma-1}$, with $k_{1},
\overline{k}_{\gamma} \neq 0$. After, we analyse the case $\mu
\neq 2$ in the presence of the same drift for some situations. We
have also discussed the connection with nonextensive statistics,
when appropriated, providing the relation between the entropic
index $q$ and the exponents appearing in the Fokker-Planck
equation. Thus, by extending here results such as those discussed
in \cite{ct2000,duxbury2,renio,tsalliseklenzi}, we hope to make
possible further applications to physical systems exhibiting
suitable mix of different forms (nonlinear $\nu \neq 1$,
fractional derivative $\mu \neq 2$ and fractal $\theta \neq 0 $)
of anomalous diffusion.

\section*{Acknowledgements}

We thank CNPq, PRONEX and Funda\c{c}\~ao Arauc\'aria (Brazilian
agencies) for partial financial support. We also thank C. Tsallis
for useful discussions.



\begin{thebibliography}{10}
\bibitem{shlesinger}
M.F. Shlesinger, G. M.  Zaslavsky and U. Frisch {\it L\'evy
Flights and Related Topics  in Physics} (Springer-Verlag, Berlin,
1994).

\bibitem{fractional}
R. Metzler and J. Klafter, Phys. Rep. {\bf 339}, 1 (2000).

\bibitem{goldenfeld}
N. Goldenfeld,{\it Lectures  On Phase Transitions  and  The
Renormalization Group}, (Addison-Wesley Publishing Company, Inc.,
Reading, 1992)

\bibitem{T88}
S.R.A. Salinas and C. Tsallis, {\it Nonextensive  Statistical
Mechanics and Thermodynamics}, Braz.  J. Phys. {\bf 29}, Number 1
(1999); S. Abe and Y. Okamoto,  {\it Nonextensive Statistical
Mechanics and Its Applications}, Lecture Notes in Physics
(Springer-Verlag, Heidelberg, 2001); P. Grigolini,  C. Tsallis and
B.J. West, {\it Classical and Quantum Complexity and Nonextensive
Thermodynamics}, Chaos , Solitons and Fractals {\bf 13}, Number 3
(Pergamon-Elsevier, Amsterdam, 2002); G. Kaniadakis, M. Lissia and
A. Rapisarda, {\it Non Extensive Thermodynamics and Physical
Applications} Physica A (Elsevier, Amsterdam, 2002), in press.

\bibitem{hilfer}
R. Hilfer, {\it Applications of Fractional Calculus in Physics}
(World Scientific, Singapore, 2000).

\bibitem{buc6}
M. Muskat, {\it The Flow of Homogeneous Fluid Through Porous Media
} (McGraw-Hill , New York, 1937).

\bibitem{buc7}
P.Y. Polubarinova-Kochina, {\it Theory of Ground Water Movement}
(Princeton University Press, Princeton, 1962).

\bibitem{buc8}
J. Buckmaster, J. Fluid Mech. {\bf 81}, 735 (1984).

\bibitem{buc10}
W.L. Kath, Physica D {\bf 12}, 375 (1984).

\bibitem{celia}
E.K. Lenzi, C. Anteneodo and L. Borland, Phys. Rev. E {\bf 63},
051109 (2001).


\bibitem{proteinfolding}
S.S. Plotkin and P.G. Wolynes, Phys. Rev. Lett. {\bf 80}, 5015
(1998).

\bibitem{relaxacao}
J. F. Douglas, in Ref. \cite{hilfer} pp. 241-331; H. Schiessel,
Chr. Friedrich and A. Blumen, in Ref. \cite{hilfer} pp. 331-376;
H. Schriessel and A. Blumen, Fractals {\bf 3}, 483 (1995).

\bibitem{anomaloustransport}
R. Metzler, E. Barkai and J. Klafter, Physica A {\bf 266}, 343
(1999).

\bibitem{prl}
B. Shaughnessy and I. Procaccia, Phys. Rev. Lett. {\bf 54}, 455
(1985).

\bibitem{correlacionada}
L. Borland, Phys. Rev. E {\bf 57}, 6634 (1998).

\bibitem{levy}
C. Tsallis, S.V.F. Levy, A.M.C. Souza  and R. Maynard, Phys. Rev.
Lett.  {\bf 75}, 3589 (1995) [Erratum: {\bf 77}, 5442 (1996)];
D.H. Zanette and P.A. Alemany, {Phys. Rev. Lett.} {\bf 75}, 366
(1995); M.O. Caceres and C.E. Budde, Phys. Rev. Lett. {\bf 77},
2589  (1996); D.H.  Zanette and P.A. Alemany, Phys. Rev. Lett.
{\bf 77}, 2590 (1996); M. Buiatti, P. Grigolini, and A.
Montagnini, Phys. Rev. Lett. {\bf 82}, 3383 (1999); D. Prato and
C. Tsallis, Phys. Rev. E {\bf 60}, 2398 (2000).



\bibitem{risken}
H. Risken, {\it The Fokker-Planck Equation} (Springer, New York,
1984).


\bibitem{Oldham}
K.B. Oldham and J. Spanier, {\it The Fractional Calculus}
(Academic Press, New York, 1974).

\bibitem{spohn}
H. Spohn, J. Phys. I France {\bf 3}, 69 (1993).

\bibitem{plastinos}
A.R. Plastino and A. Plastino, Physica A {\bf 222}, 347 (1995); C.
Giordano, A.R. Plastino, M. Casas and A. Plastino, Eur. Phys. J. B
{\bf 22}, 361 (2001).

\bibitem{buckman}
C. Tsallis and D.J. Bukman, Phys. Rev. E {\bf 54}, R2197 (1996).

\bibitem{isabel}
I. T. Pedron , R. S. Mendes,  L. C. Malacarne and E. K. Lenzi,
Physical Review E {\bf 65}, 41108 (2002).

\bibitem{ct2000}
M. Bologna, C. Tsallis and P. Grigolini, Phys. Rev. E {\bf 62},
2213 (2000).

\bibitem{duxbury2}
C. Tsallis, Physica A {\bf 221}, 227 (1995).


\bibitem{renio}
L.C. Malacarne, I.T. Pedron, R.S. Mendes and E.K. Lenzi, Phys.
Rev. E {\bf 63}, 30101R (2001).

\bibitem{tsallisnonextensive}
C. Tsallis, J. Stat. Phys. {\bf 52}, 479 (1988); E.M.F. Curado and
C. Tsallis, J. Phys. A {\bf 24}, L69 (1991) [Corrigenda: {\bf 24},
3187 (1991) and {\bf 25}, 1019 (1992)];  C. Tsallis, R.S. Mendes
and A.R. Plastino, Physica A {\bf 261}, 534 (1998).

\bibitem{tsalliseklenzi}
C. Tsallis and E. K. Lenzi, Chem. Phys. in press (2002).

\bibitem{strato}
C. W. Gardiner, {\it Handbook of Stochastic Methods: For Physics, Chemistry
and the Natural Sciences}, (Springer Series in Synergetics, New York, 1996).
\end{thebibliography}
\end{document}